\documentclass[runningheads]{svmult}

\usepackage{makeidx}   
\usepackage{graphicx}  
\usepackage{subeqnar}  
\usepackage{multicol}  
\usepackage{physprbb}  
\makeindex             

\newcommand{\greeksym}[1]{{\usefont{U}{psy}{m}{n}#1}}

\newcommand{\uDelta}{\mbox{\greeksym{D}}}

\begin{document}
\title*{Some Mathematical And Numerical Questions Connected With First And
  Second Order Time Dependent Systems Of Partial Differential Equations}
\toctitle{Some Questions Connected With Systems Of PDE}
%
%
\titlerunning{Mathematical And Numerical Questions On Systems Of PDE}
%
\author{Heinz-O. Kreiss\inst{1}
\and Omar E. Ortiz\inst{2}}
\authorrunning{Heinz-O. Kreiss and Omar E. Ortiz}
%
%
\institute{Department of Mathematics, University of California, Los
Angeles, CA 90095, USA \and Universidad Nacional de C\'ordoba,
Facultad de Matem\'atica, Astronom\'{\i}a y F\'{\i}sica, (5000)
C\'ordoba, Argentina.}

\maketitle              

\section{Introduction}
There is a tendency to write the equations of general relativity as a first
order symmetric system of time dependent partial differential equations.
However, for numerical reasons, it might be advantageous to use a second order
formulation like one obtained from the ADM equations. Unfortunately, the type
of the ADM equations is not well understood and therefore we shall discuss, in
the next section, the concept of wellposedness. We have to distinguish between
weakly and strongly hyperbolic systems. Strongly hyperbolic systems are well
behaved even if we add lower order terms. In contrast; for every weakly
hyperbolic system we can find lower order terms which make the problem totally
illposed.  Thus, for weakly hyperbolic systems, there is only a restricted
class of lower order perturbations which do not destroy the wellposedness. To
identify that class can be very difficult, especially for nonlinear
perturbations.  In Section 3 we will show that the ADM equations, linearized
around flat with constant lapse function and shift vector, are only weakly
hyperbolic. However, we can use the trace of the metric as a lapse function to
make the equations into a strongly second order hyperbolic system.

Using simple models we shall, in Section 4, demonstrate that approximations of
second order equations have better accuracy properties than the corresponding
approximations of first order equations. Also, we avoid spurious waves which
travel against the characteristic direction.

In the last section we discuss some difficulties connected with the
preservation of constraints.

\section{Well Posed Problems}
\subsection{First Order Systems} 
Consider the Cauchy problem for a first order system with constant
coefficients
\begin{eqnarray}
&&\vec{u}_t = \sum_{j=1}^s A_j D_j \vec{u} =: P(D) \vec{u},
\hspace{1cm} D_j=\frac{\partial}{\partial x_j}, \label{1.1}\\
&&\vec{u}(t=0) = \vec{f}(x),
\hspace{1cm} x=(x_1, \dots, x_s), \; -\infty < x_j < \infty.
\nonumber
\end{eqnarray}
We construct simple wave solutions
\[
\vec{u}(x,t)= e^{\I \langle\omega,x\rangle} \widehat{\vec{u}}(\omega, t),
\;\;\; \omega= (\omega_1, \dots, \omega_s) \; \mbox{ real}
\]
and obtain
\begin{equation}
\widehat{\vec{u}}_t= \I |\omega| \sum_{j=1}^s A_j \omega'_j
\widehat{\vec{u}} =: \I |\omega| P(\omega')
\widehat{\vec{u}}, \hspace{1cm} \omega'=
\omega/|\omega| \label{1.2}
\end{equation}

\begin{definition} We call the problem (\ref{1.1}) strongly hyperbolic 
  if for every $\omega'$ the eigenvalues of the symbol $P(\omega')$ are real
  and there is a complete set of uniformly (in $\omega'$) linearly independent
  eigenvectors.
\end{definition}

Examples of strongly hyperbolic systems are symmetric systems where
$A_j=A_j^\dagger$. 

The solutions of strongly hyperbolic systems satisfy an energy estimate
\begin{equation}
\|\vec{u}(\cdot,t)\|^2 \le K^2 \E^{2\alpha t}
\|\vec{u}(\cdot,0)\|^2. \label{1.3}
\end{equation}
Here $K, \alpha$ are universal constants which do not depend on the initial
data $\vec{u}(x,0)=\vec{f}(x)$. The norms are $L_2$ norms. For systems
(\ref{1.1}) with constant coefficients $\alpha = 0$.

Strong hyperbolicity and the existence of an energy estimate are equivalent,
we have\footnote{First order theory is well known, we refer to \cite{kl}.}
\begin{theorem}
  The solutions of $(\ref{1.1})$ satisfy an energy estimate of the type
  $(\ref{1.3})$ if and only if the system is strongly hyperbolic.
\end{theorem}

The most important property of strongly hyperbolic systems is that we can add
lower order terms and an estimate of type (\ref{1.3}) is still valid. We have
\begin{theorem}
  Let $(\ref{1.1})$ be strongly hyperbolic. Then the solutions of
\begin{equation}
\vec{w}_t= P(D)\vec{w} + B\vec{w}
\end{equation}
satisfy an estimate of type $(\ref{1.3})$. Here $B$ is any bounded operator.
\end{theorem}

\begin{definition}
  We call the problem $(\ref{1.1})$ weakly hyperbolic if the eigenvalues of
  $P(\omega')$ are real.
\end{definition}
In this definition we do not require that there is a complete set of
eigenvectors. An example in dimension one is given by
\[
\vec{u}_t= \left(\begin{array}{cc}
1 & 1\\
0 & 1
\end{array}\right) \vec{u}_x =: A \vec{u}_x
\]
Simple wave solutions for this system have the form
\[
\vec{u}(x,t)= \E^{\I\omega A t} \E^{\I \omega x} \widehat{\vec{u}}(\omega,0)=
\left(I + \I\,\omega\left(\begin{array}{cc}
      1 & 1\\
      0 & 1
\end{array}\right) t\right) \E^{\I\omega(x+t)} \widehat{\vec{u}}(\omega,0)
\]
Thus there is no exponential growth but there is the polynomial growth in
$\omega t$. This is typically the case for weakly hyperbolic systems. One can
prove
\begin{theorem}
  For weakly hyperbolic systems the growth of simple wave solutions is at most
  of the order ${\cal O}(1+|\omega t|^{n-1})$, where $n$ is the number of
  components of $\vec{u}$.
\end{theorem}

The real difficulty with weakly hyperbolic systems is that lower order
terms can make them exponentially ill posed. For example, consider
\[
\vec{u}_t= \left(\begin{array}{cc}
1 & 1\\
0 & 1
\end{array}\right) \vec{u}_x + \left(\begin{array}{cc}
0 & 0\\
1 & 0
\end{array}\right) \vec{u}.
\]
Making the simple wave ansatz
\[
\vec{u}(x,t)= \E^{\I \omega x} \widehat{\vec u}(\omega,t)
\]
we obtain
\[
\widehat{\vec{u}}_t = \left(\begin{array}{cc}
\I \omega & \I \omega \\
1 & \I \omega
\end{array}\right) \widehat{\vec{u}} =: A \widehat{\vec{u}}.
\]
The eigenvalues $\lambda$ of $A$ are given by
\[
\lambda= \I \omega \pm \sqrt{\I \omega}, \;\;\; \mbox{i.e. Re} \lambda = \pm
\frac{\sqrt{2}}{2} \sqrt{|\omega|}.
\]
Therefore the perturbed problem is exponentially ill posed.

The same result holds if we consider the variable coefficient problem
\[
\vec{u}_t= U^\dagger (x)\left(\begin{array}{cc}
1 & 1\\
0 & 1
\end{array}\right) U(x) \vec{u}_x, \;\;\; U(x)= \left(\begin{array}{cc}
\cos x & \sin x \\
\sin x & \cos x
\end{array}\right).
\]

There are no problems to generalize the results to variable coefficients and
quasi-linear systems if the system is pointwise strongly hyperbolic.

Unfortunately, in applications one can be confronted with systems which are
weakly hyperbolic. In this case one has to carefully study the influence of
lower order terms. For example, trivially,
\[
u_t + (u^2)_x + (v^2)_x = -\alpha u, \;\;\; v_t+ (v^2)_x = -\alpha v
\]
is well behaved ($\alpha >0$ sufficiently large so that no shocks form). We
can solve the second equation to obtain $v$ which becomes a governing
function in the first equation.

\subsection{Second Order Systems}
We consider second order systems
\begin{equation}
\vec{u}_{tt}= P_0(D) \vec{u} + P_1(D) \vec{u}_t \label{1.5}
\end{equation}
where
\[
P_0(D) = \sum_{j,k=1}^s A_{jk} D_j D_k, \;\;\; P_1(D) = \sum_{j=1}^s A_j D_j.
\]
We calculate simple wave solutions. Introducing
\[
\vec{u}(x,t)= e^{\I \langle\omega,x\rangle} \widehat{\vec{u}}(\omega, t)
\]
gives us
\begin{equation}
\widehat{\vec{u}}_{tt}= -|\omega|^2 P_0(\omega') \widehat{\vec{u}} + \I
|\omega| P_1(\omega') \widehat{\vec{u}}_t. \label{1.6}
\end{equation}

We have
\begin{lemma}
A necessary condition for well posedness is that, for all $\omega'$, the
eigenvalues $\tilde{\kappa}$ of
\begin{equation}
[-\tilde{\kappa}^2 I + P_1(\omega') \tilde{\kappa} + P_0(\omega')] 
\vec{a} = 0 \label{1.7}
\end{equation}
are real.
\end{lemma}
\begin{proof}
  If $\tilde\kappa(\omega'), \vec{a}(\omega')$ is a solution of (\ref{1.7})
  then $-\tilde\kappa(\omega'), \vec{a}(\omega')$ is a solution if we
  replace $\omega'$ by $-\omega'$. Since the solutions of (\ref{1.6}) are of
  the form $e^{\I|\omega|\tilde\kappa(\omega') t} \vec a(\omega')$ we only
  avoid catastrophic growth if $\tilde\kappa$ is real.
\end{proof}
If $P_1 =0$ then (\ref{1.7}) becomes
\[
[-\tilde\kappa^2 I+ P_0(\omega')]\vec{a}= 0
\]
and Lemma 1 reduces to
\begin{lemma}
If $P_1=0$ then a necessary condition for well posedness is that the
eigenvalues of $P_0(\omega')$ are real and nonnegative.
\end{lemma}

We can write (\ref{1.6}) as a first order system by introducing a new variable
$\widehat{\vec{v}}$ with $\widehat{\vec{u}}_t= \I |\omega| \widehat{\vec{v}}$.
We obtain
\begin{equation}
\left(\begin{array}{c}
\widehat{\vec{v}} \\ \widehat{\vec{u}} 
\end{array} \right)_t = \I |\omega| \left(\begin{array}{cc}
P_1(\omega') & P_0(\omega') \\
I & 0
\end{array}\right) \left(\begin{array}{c}
\widehat{\vec{v}} \\ \widehat{\vec{u}} 
\end{array} \right) =: \I |\omega| \widehat{\bf P} \left(\begin{array}{c}
\widehat{\vec{v}} \\ \widehat{\vec{u}} 
\end{array} \right) \label{1.8}
\end{equation}
The eigenvalues of $\widehat{\bf P}$ are determined by (\ref{1.7}). Using this
reduction, we can define what we mean by strongly and weakly hyperbolic
(second order) systems.
\begin{definition}
  The system $(\ref{1.5})$ is called strongly hyperbolic if for all $\omega'$
  the eigenvalues of $\widehat{\bf P}$ are real and there is a uniformly
  linearly independent (in $\omega'$) complete set of eigenvectors.
\end{definition}

For strongly hyperbolic systems one can again develop a rather complete
theory for local existence of quasi-linear systems. In particular lower order
terms
\[
Q \vec{u} = \sum_{j=1}^s B_j D_j \vec{u} + B_0 \vec{u}_t + C \vec{u}
\]
do not destroy the well posedness of the problem.

If $P_1 \equiv 0$ we have
\begin{theorem}
  Assume that $P_1 \equiv 0$. The system is strongly hyperbolic if and only if
  the eigenvalues of $P_0(\omega')$ are strictly positive and $P_0(\omega')$
  has a complete set of eigenvectors which is uniformly (in $\omega'$)
  linearly independent.
\end{theorem}
\begin{proof}
  Notice that when $P_1 \equiv 0$, any eigenvector of $\widehat{\bf P}$ with
  eigenvalue $\lambda_j(\omega')$ is of the form
\begin{equation}
\left(\begin{array}{c} \lambda_j \vec a_j \\ \vec a_j \end{array} \right)
\label{eigvec}
\end{equation}
where the splitting corresponds to the block structure of $\widehat{\bf P}$.
Moreover, for each eigenvector of the form (\ref{eigvec}) there is another
eigenvector
\[
\left(\begin{array}{c} -\lambda_j \vec a_j \\ \vec a_j \end{array} \right)
\]
with eigenvalue $-\lambda_j(\omega')$, which is linearly independent from the
first if and only if $\lambda_j \neq 0$.

Now, it is easy to check that a set of eigenvectors
\[
\left\{\left(\begin{array}{c} \lambda_j \vec a_j \\ \vec a_j \end{array}
  \right); \left(\begin{array}{c} -\lambda_j \vec a_j \\ \vec a_j \end{array}
  \right)\right\}, \qquad j= 1,2, \dots n.
\]
with real eigenvalues $\{\lambda_j, -\lambda_j\}, \; j=1,2, \dots,n$ is a set
of $2n$ uniformly linearly independent (in $\omega'$) eigenvectors if and only
if the set $\{\vec a_j(\omega')\},$ $j=1,2,\dots n,$ is a set of uniformly
linearly independent (in $\omega'$) eigenvectors of $P_0(\omega')$ with
positive eigenvalues $\lambda_j^2(\omega') >0$. This proves the theorem.
\end{proof}
\begin{definition}
  The system $(\ref{1.5})$ is called weakly hyperbolic if for all $\omega'$
  the eigenvalues of $\widehat{\bf P}$ are real.
\end{definition}
In particular we have
\begin{lemma}
  If $P_1 \equiv 0$ and $P_0(\omega')$ has zero as an eigenvalue then the
  system is not strongly hyperbolic. It is weakly hyperbolic if the
  eigenvalues are real and non negative.
\end{lemma}

As in the case of first order systems, weakly hyperbolic systems can have
catastrophic exponential growth when adding lower order terms or considering
variable coefficients. As example we consider
\[
u_{tt}=au_{xx}+u_{yy}+bu_x+cu_y
\]
The problem is strongly hyperbolic if $a>0$ and weakly hyperbolic if $a=0$ and
there is catastrophic exponential growth if $a=0$ and $b \neq 0.$

In the next section we will show that the ADM equations, linearized around
flat, are only weakly hyperbolic for constant lapse function and shift vector.
We can transform it into a strongly hyperbolic system if we choose the lapse
function proportional to the trace of the solution. However, such a choice
might introduce singularities.

Consider, for example,
\[
u_{tt}=\alpha_xu_x+\alpha_{xx}.
\]
If $\alpha=\alpha(x,t)$ is a given function, then the equation is weakly
hyperbolic. If we choose $\alpha=u,$ we obtain
\[
u_{tt}=(u_x)^2+u_{xx}.
\]
Now the equation is strongly hyperbolic but we will, in general, encounter
singularities due to the lower order term.

\section{Second Order Initial Value Formulations For General Relativity}
Our starting point are the ADM equations \cite{adm} for General Relativity.
The 3-metric induced on the spacelike 3-surfaces $t=constant$ is denoted by
$\gamma_{ij},$; all latin indices run over $1,2,3$.

From start we fix the shift vector equal to zero but keep a general lapse
function $\alpha$.  Using the ADM equation for $\gamma_{ij}$ to eliminate the
extrinsic curvature from the other ADM equation we get a second order
evolution equation for $\gamma_{ij}$
\begin{eqnarray}
\partial_t^2 \gamma_{ij}&=& \alpha^2 \gamma^{lm} \Bigl(\partial_l\partial_m
\gamma_{ij} + \partial_i\partial_j \gamma_{lm} -
\partial_i\partial_l\gamma_{mj} - \partial_j\partial_l\gamma_{mi}\Bigr) + 2
\alpha \partial_i\partial_j \alpha \nonumber \\
&& + \; {\rm lot}, \label{adm2nd}
\end{eqnarray}
where all derivatives are partial derivatives with respect to time and the
coordinates on the $t= constant$ 3-surfaces. Here and below ``lot'' stands for
``lower order terms'', that is functions of $\alpha$, $\gamma_{ij}$ and their
first derivatives. For the purpose of this paper, it is enough to say that all
lower order terms are quadratic in first order derivatives.

We have to consider the constraint equations. The momentun constraint is
\[
\gamma^{jl} \partial_t \Bigl(\partial_i \gamma_{jl} - \partial_j
\gamma_{il}\Bigr) + {\rm lot} =0,
\]
while the Hamiltonian constraint is 
\[
\gamma^{ij} \gamma^{lm} \Bigl(\partial_i \partial_l \gamma_{jm} - \partial_i
\partial_j \gamma_{lm}\Bigr) + {\rm lot} =0.
\]

We now linearize around a flat solution (Minkowski spacetime) in Cartesian
coordinates, that is we make
\begin{equation}
\gamma_{ij} = \delta_{ij} + \varepsilon \; h_{ij} \qquad \mbox { and } \qquad 
\gamma^{lm} = \delta^{lm} - \varepsilon \; \delta^{lp} \delta^{mq} h_{pq} 
+ {\cal O}(\varepsilon^2)
\end{equation}
and keep terms linear in $\varepsilon$. Our new variable is $h_{ij}$. The
constraint equations become
\[
\partial_t \Bigl(\partial_i H - \delta^{lm} \partial_l h_{im} \Bigr)= 0
\]
and
\[
\delta^{ij} \delta^{lm} \partial_i \partial_l h_{im} - \uDelta H= 0
\]
where $H= {\rm tr }(h_{ij})= \delta^{ij} h_{ij}$ and $\uDelta$ is the
Laplacian in ${\bf R}^3$. Both linearized constraint equations are satisfied
if
\begin{equation}
\partial_i H - \delta^{lm} \partial_l h_{im}= 0. \label{constraint}
\end{equation}

Before linearizing (\ref{adm2nd}) we have to make a choice of lapse. On the
one hand the simplest possible choice $\alpha\equiv 1$ gives, after
linearization,
\begin{eqnarray*}
\partial_t^2 h_{ij} &=& \delta^{lm} \Bigl(\partial_l\partial_m
h_{ij} + \partial_i\partial_j h_{lm} -
\partial_i\partial_l h_{mj} - \partial_j\partial_l h_{mi}\Bigr) \\
&=& \uDelta h_{ij} + \partial_j \Bigl(\partial_i H -\delta^{lm} 
\partial_l h_{mi}\Bigr) - \partial_i\partial_j H 
\end{eqnarray*}
so that (\ref{constraint}) implies
\begin{equation}
\partial_t^2 h_{ij} = \uDelta h_{ij} - \partial_i\partial_j H.  \label{linear1}
\end{equation}
On the other hand, choosing
\[
\alpha= \frac{k}{3}\mbox{tr}(\gamma_{ij})= \frac{k}{3} \delta^{lm}
\gamma_{lm},
\]
gives after linearization and using (\ref{constraint})
\begin{equation}
\partial_t^2 h_{ij} = k^2 \Bigl(\uDelta h_{ij} - \frac{1}{3} \partial_i 
\partial_j H\Bigr).  \label{linear2}
\end{equation}

We define $\vec{u}= (h_{11}, h_{22}, h_{33}, h_{12}, h_{13}, h_{23})^{\rm t}$
to analyze the hyperbolicity of equations (\ref{linear2}) and (\ref{linear1}).

Thus the matrix $P_0(\omega')$, as defined in Section 2.2, of the system
asociated to (\ref{linear2}) has constant eigenvalues: $k^2$ with multiplicity
five, and $(2/3) k^2$ with multiplicity one. Also, this matrix has a uniformly
linearly independent complete system of eigenvectors. Then according to
Theorem 4, the system is strongly hyperbolic for any $k \neq 0$. $|k| \leq 1$
gives a system with characteristic speeds smaller or equal than one, while
$|k| > 1$ would be an ``unphysical'' (though strongly hyperbolic) system with
characteristic speeds higher than one.

The matrix $P_0(\omega')$ of the system asociated to (\ref{linear1}) is
uniformly diagonable but its eigenvalues are: $0$ with multiplicity one and
$1$ with multiplicity five. Thus, according to Lemma 3, equation
(\ref{linear1}) is only weakly hyperbolic.

Another possibility to ``cure'' equation (\ref{linear1}) is the usual choice
of coupling the lapse function to the determinant of the 3-metric instead of
coupling it to the trace as we have done above. This also leads to a strongly
hyperbolic system.

\section{Difference Approximations}
Consider the simple model problem
\begin{equation}
u_t=u_x,\quad -\infty <x<\infty,~ t\ge 0 \label{4.1}
\end{equation}
with initial data
\begin{equation}
u(x,0)=f(x). \label{4.2}
\end{equation}
We are interested in solutions which are $2\pi$-periodic in space. We want to
solve the above problem by difference approximation. For that reason we
introduce a gridlength $h=2\pi /N, ~ N$ a natural number, gridpoints
$x_\nu=\nu h$ and gridfunctions $ u_\nu(t)=u(x_\nu,t).$ We also introduce the
usual centered difference operators by

\begin{eqnarray*}
\partial/\partial x &\sim& D_0u_\nu = (u_{\nu+1}-u_{\nu-1})/2h, \\
\partial^2/\partial x^2 &\sim& D_+D_- u_\nu= (u_{\nu+1}-2u_\nu+u_{\nu-1})/h^2.
\end{eqnarray*}
Then we approximate (\ref{4.1}),(\ref{4.2}) by
\begin{equation}
(\tilde u_\nu)_t=D_0\tilde u_\nu,\quad \nu=0,1,2,\ldots,N-1
\label{4.3}
\end{equation}
with periodic boundary conditions
\begin{equation}
\tilde u_\nu(t)=\tilde u_{\nu+N}(t) \label{4.4}
\end{equation}
and initial conditions
\begin{equation}
\tilde u_\nu(0)=f_\nu. \label{4.5}
\end{equation}
(\ref{4.3}) -- (\ref{4.5}) represents a system of ordinary differential
equations which we solve with help of a standard ODE solver like the usual
Runge-Kutta method.

We want to discuss the accuracy of the approximation. We assume that
\[
f(x)=\sum_{\omega=-M}^M e^{i\omega x} \hat f(\omega ),\quad M\le N/2.
\]
Then we can expand both the solutions of (\ref{4.1}),(\ref{4.2}) and
(\ref{4.3})-(\ref{4.5}) into Fourier polynomials
\begin{equation}
u(x,t)=\sum_{\omega=-M}^M e^{i\omega x} \hat u(\omega,t),\quad
\tilde u_\nu(t)=\sum_{\omega=-M}^M e^{i\omega x_\nu} \hat{\tilde u}(\omega,t)
\label{4.6}
\end{equation}
with
\[
\hat u(\omega,0)=\hat{\tilde u}(\omega,0)=\hat f(\omega ).
\]
We introduce (\ref{4.6}) into (\ref{4.1}) and (\ref{4.3}), respectively. Since
\[
\partial e^{i\omega x} /\partial x=i\omega e^{i\omega x} \quad {\rm and}\quad
D_0e^{i\omega x}={i\sin \omega h\over h} e^{i\omega x},
\]
we obtain, for every frequency,
\[
\hat u_t (\omega,t)=i\omega\hat u (\omega,t),\quad
\hat{\tilde u}_t (\omega,t)=i\alpha\omega\hat{\tilde u}(\omega,t),~
\alpha={\sin \omega h\over \omega h}.
\]
Therefore,
\[
\hat u(\omega,t)=e^{i\omega t}\hat f(\omega ),\quad
\hat{\tilde u}(\omega,t)=e^{i\alpha \omega t}\hat f(\omega ).
\]
Thus there is a phase error
\[
e=(1-\alpha) \omega t.
\]
Also, the wave speed for the difference approximation depends on the
frequency, i.e., there is dispersion for the difference approximation but not
for the differential equation. This causes lots of difficulties if the
solution is not properly resolved, i.e., there are not enough
points/wavelength.

Instead of the above second order method, one can use higher order methods.
This results in a remarkable improvement of the accuracy. In Table
\ref{table1} we give the number of points/wavelength so that the numeric
solution has a phase-error of 10\% or 1\% after calculating during $q$ time
periods with methods of different order.
\begin{table}
\caption{Points/wavelength}
\begin{center}
\renewcommand{\arraystretch}{1.4}
\setlength\tabcolsep{0pt}
\begin{tabular}{cccc}
\hline\noalign{\smallskip} 
\hspace{5mm} $e$ \hspace{5mm} & \hspace{5mm}{\rm 2nd Order}\hspace{5mm}
& \hspace{5mm}{\rm 4th Order}\hspace{5mm} & \hspace{5mm}{\rm 6th
  Order}\hspace{5mm} \\ \hline 
10\% & 20 $q^{1/2}$ & 7 $q^{1/4}$ & 5 $q^{1/6}$ \\ 
1\% & 64 $q^{1/2}$ & 13 $q^{1/4}$ & 8 $q^{1/6}$ \\ 
\hline
\end{tabular}
\end{center}
\label{table1}
\end{table}
\goodbreak P. Huebner \cite{h} and J. Thornburg \cite{t}, using fourth order
accurate methods, have demonstrated the improved accuracy for the Einstein
equations.

If we calculate with $N$ points, then the solution consists of $M\sim N/2$
simple waves. Most of them have large phase errors.  Therefore, the
approximation is only useful if the Fourier expansion of the analytic solution
converges rapidly. In particular, the part of the solution consisting of those
waves with few points/wavelength travels in the wrong direction.

Consider
\[
\hat{\tilde  u}_{\nu t}=D_0\hat{\tilde u}_\nu
\]
with highly oscillatory initial data
\[
\hat{\tilde u}_\nu(0)=(-1)^\nu g_\nu,\quad g~{\rm smooth~function}.
\]
Introduce a new variable by $\tilde u_\nu=(-1)^\nu w_\nu.$ Then $w$ solves
\begin{eqnarray*}
w_{\nu t}&=&-D_0 w_\nu,\\
w_\nu(0)&=&g_\nu,
\end{eqnarray*}
which approximates
\begin{eqnarray*}
w_t&=&- w_x,\\
w(x,0)&=&g(x).
\end{eqnarray*}
Thus $\hat{\tilde u}_\nu(t)$ represents a highly oscillatory wave which
travels in the ``wrong'' direction. The usual way to control $\hat{\tilde
  u}_\nu(t)$ is to add an artificial viscosity term.

As an example, consider
\[
u_t=x\, u_x,\quad -1\le x\le 1,
\]
with boundary condition
\[
u(-1,t)=-1,\quad u(1,t)=1,
\]
and the initial data
\[
u(x,0)=-\cos {\pi\over 2} (x+1).
\]
The solution forms an internal layer at $x=0$ where the gradient becomes larger
and larger. If we use the approximation
\[
\tilde u_{\nu t}=x_\nu D_0\tilde u_\nu,
\]
then there will be a highly oscillatory wave traveling out of the layer.

We approximate the wave equation
\[
u_{tt}=u_{xx}
\]
by
\[
\tilde u_{tt}=D_+D_-\tilde u.
\]
For the same level of accuracy, we need only half the number of 
points/wave\-length. Also, there are no spurious waves which travel in
the wrong direction.

\section{Constraints}
Using an example from fluid dynamics we want to demonstrate some of the
problems which can arise when solving equations with constraints.
Consider the Stokes problem
\begin{eqnarray}
u_t+p_x &=&\uDelta u, \label{5.1}\\
v_t+p_y &=&\uDelta v, \label{5.2}\\
d=:u_x+v_y &=&0, \label{5.3}
\end{eqnarray}
in some domain $\Omega\times (0,T)$ with boundary conditions
\[
u_n=0\quad {\rm for}\quad (x,y)\in\partial\Omega,\quad\quad  0\le t\le T.
\]
Here $u,v$ denote the velocity components, $u_n$ the normal component
and $p$ the pressure. Differentiating the first equation with
respect to $x$ and the second with respect to $y$ gives us, using
the divergence relation $d=0,$
\begin{equation}
\uDelta p=0. \label{5.4}
\end{equation}
We solve (\ref{5.1}),(\ref{5.2}) and (\ref{5.4}) and think of (\ref{5.3}) as
the constraint.

One could be tempted to use, for $p,$
\begin{equation}
p=p_0,\quad (x,y)\in\partial \Omega,\quad p_0~{\rm given~function,}
\label{5.5}
\end{equation}
as boundary condition. However, then we would, in general, not preserve 
the constraint $ d=0.$

Differentiating (\ref{5.1}) with respect to $x$ and (\ref{5.2}) with respect
to $y$ and using (\ref{5.4}) gives us an equation for the divergence $d,$
\[
d_t=\uDelta d. 
\]
By assumption, $d=0$ for $t=0$ but we cannot guarantee that $d=0$
at later times if we use the boundary conditions (\ref{5.5}). We must use
\[
d=0\quad {\rm for}\quad (x,y)\in\partial \Omega,~ t\ge 0,
\]
as boundary condition and we cannot give $p.$

Let $\uDelta_h=D_{+x}D_{-x}+D_{+y}D_{-y}.$
A typical difference approximation is given by
\begin{eqnarray}
\tilde u_t+D_{0x}\tilde p &=&\uDelta_h \tilde u, \label{5.6}\\
\tilde v_t+D_{0y}\tilde p &=&\uDelta_h \tilde v, \label{5.7}\\
\uDelta_h \tilde p &=&0. \label{5.8}
\end{eqnarray}
For the discrete divergence $d_h=D_{0x}\tilde u+D_{0y}\tilde v$ we then
obtain
\begin{eqnarray}
d_{ht}&=&-(D^2_{0x}+D^2_{0y})\tilde p+\uDelta_h d_h, \label{5.9}\\
\uDelta_h\tilde p&=&0. \label{5.10}
\end{eqnarray}
The difficulty here is that
\[
\uDelta_h\ne D^2_{0x}+D^2_{0y}
\]
and therefore divergence is created. Instead of (\ref{5.8}) one can use
\[
\uDelta_h\tilde p=\alpha d_h,\quad \alpha >\! > 1~{\rm constant.}
\]
Then we can write (\ref{5.9}) as
\[
d_{ht}+(D^2_{0x}+D^2_{0y}-\uDelta_h)\tilde p+\alpha d_h=\uDelta_hd_h.
\]
The damping term $\alpha d_h$ keeps $d_h$ under control.

\end{document}